# Simultaneous Switching of Multiple GaN Transistors in a High-Speed Switch

Daniil Frolov and Greg Saewert
Fermi National Accelerator Laboratory
Batavia, IL 60510
dfrolov@fnal.gov

*Abstract*— A broadband travelling wave kicker operating with 80 MHz repetition rates is required for the new PIP-II accelerator at Fermilab. We present a technique to drive simultaneously four series-connected enhancement mode GaN-on-silicon power transistors by means of microwave photonics techniques. These four transistors are arranged into a high voltage and high repetition rate switch. Using multiple transistors in series is required to share switching losses. Using a photonic signal distribution system is required to achieve precise synchronization between transistors. We demonstrate 600 V arbitrary pulse generation into a 200 Ohm load with 2 ns rise/fall time. The arbitrary pulse widths can be adjusted from 4 ns to essentially DC.

Keywords — PIP-II, GaN, fast switch, beam kicker, gate synchronization

## I. Introduction

Future linear accelerators operating CW will require ever faster CW-compatible traveling wave kickers for bunch-by-bunch selection. For example, the unique feature of the PIP-II machine is the capability of delivering proton beam to several experiments simultaneously with a beam structure that can be adjusted to each experiment's needs [1]. Bunch-by-bunch selection is done by kicking beam bunches either out of the machine or allowing them to pass with a kicker composed of a beam deflector and its corresponding voltage driver [2]. The deflector is a broadband dual-helix traveling wave structure. The voltage driver, described here, is designed as a fast, relatively high voltage, switch. Each helix is driven with its own driver biased with opposite polarity voltage. The operational requirements of the PIP-II linac warm front end kicker have greater than 500 V on each helix, an absolute worst case 4.0 ns rise/fall time (5-95%) and switching rates of 45 MHz for 0.55 ms bursts at 20 Hz intervals. When the PIP-II eventually operates CW, overall average switching rates will extend from one half to 30 MHz. Moreover, the machine's capability to deliver unique beam structure to each experiment requires arbitrary pulse width and duty factor generation.

Such requirements represent a significant leap forward from what has been achieved with existing technology of high repetition rate, high voltage pulse generators. At present, there are no devices available on the market that can satisfy these requirements. Therefore, intensive R&D work has been done at Fermilab during last six years to advance fast switching technology required for the PIP-II accelerator.

In recent years, studies of wide-bandgap semiconductors (silicon carbide, gallium nitride, aluminum nitride, etc.) and devices based on them have become very active. The unique properties of these semiconductor materials (large bandgap, high mobility of charge carriers and their saturation rates, large thermal conductivity coefficients, etc.) are attractive for devices with record values of power, voltage and current. The most promising wide-bandgap material now is GaN. The record specific power density is one of the most outstanding properties of modern GaN transistors. The maximum critical electric field strength (10 times greater than that of silicon) makes it possible to achieve breakdown voltages of 1-1.5 kV and raise the operating voltage of the drain to 500-700 V, which, in combination with the high current density, provides a specific output power of industrial GaN transistors in the range of 10-40 W per 1 millimeter of the gate width, which is an order of magnitude greater than the specific output power of GaAs transistors. Although modern GaN transistors are at least a factor of five times both lower in capacitance and faster in switching speed than MOSFETs, for a single GaN transistor, switching rates of tens of MHz will have unacceptable switching losses, if it were to switch at 500 V. However, losses per transistor can be reduced by using multiple transistors in series as one switch; and, switching losses are shared. The design task is to assure they switch on and off simultaneously to share voltage and power loss.

## II. The Kicker

The PIP-II travelling wave deflector is shown in Fig. 1. It comprises two helical travelling wave structures that are used to slow down the electromagnetic wave propagation velocity to match the beam velocity [3, 4]. These structures are driven separately, each with its own driver. In the PIP-II medium energy beam transport section, the bunch length is approximately 1.3 ns, and the beam is bunched at 6.15 ns intervals. The system can kick out a beam on a bunch-by-bunch basis. Thus, each helix and its driver have a rise/fall time of 2 ns and a minimum pulse width of 4 ns. This is necessary to ensure the system can turn on and off completely in between the bunches. Voltages can be generated on and off to kick and pass bunches having pulse widths that can be from 4 ns to essentially DC. Any arbitrary pulse pattern can be generated and therefore any beam structure can be achieved.

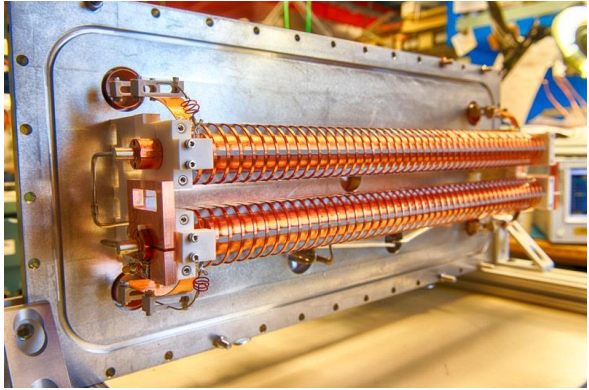

Fig. 1. The PIP-II dual-helix travelling wave beam deflector

### III. THE HIGH SPEED GATE DRIVER

Each helix of the kicker is driven with the individual switch biased to + 500 V or – 500 V as shown in fig. 2. Each switch consists of four GaN-on-silicon enhancement mode transistors GS66502B from GaN Systems Inc [5]. Two major problems were solved during the development of the of these high-speed switches.

The first problem is related to the gate drivers. Currently there are no available FET gate drivers on the market that can deliver parameters required for our application. These parameters are: 5 V, 300 mA, <2 ns rise/fall time, 4 ns minimum pulse width with up to 100 MHz repetition rate. We have developed such gate driver which is shown in fig. 3. An AC voltage source at the ground level delivers power to each driver circuit via transformer T1. This AC voltage is rectified and converted to DC voltage from which three different well-regulated DC voltages are produced in the section labeled "isolated DC power supplies". Transformer T1 is designed to hold off at least 600 V, deliver more than 2 Watts of power to the isolated gate driver circuit and present only several picofarads of capacitance across the isolation. The capacitance is made low by delivering AC power at 500 kHz that allows the use of a physically small ferrite core transformer. Minimizing capacitance between each gate driver circuit and ground is necessary for the multi-FET switch assembly to switch completely on and off in less than 2 nanoseconds.

The light signal is received by the photo detector D1, converted to an electrical signal and amplified in the stage labeled "photo detection and gain". Depending on the configuration of the multi-FET switch, the reception of light will either define the time the GaN FET is to be on and no light will define the time the GaN FET will be off – or vice versa. The polarity of this operation is set by making a wire connection in the signal gain stage. All FETs comprising a single switch assembly are configured the same way.

The "leading and trailing edge delay adjustment" section provides timing adjustment. Both the turn-on and turn-off edges are adjusted and set individually for all the FETs in a switch to be matched to less than 0.1 nanoseconds. The output of this section has the current drive capability to drive transistor Q1 in the "GaN FET gate driver" section.

The arrangement of parts and choice of transistors of this GaN FET gate driver section is novel. The relatively low power consumption of this circuit keeps the physical size of the entire driver circuit small enough to minimize parasitic capacitance to ground and not degrade the overall switch performance.

Operation of the GaN FET gate driver section is as follows. When voltage to the gate of Q1 is applied to turn it on, its drain goes low that drives the gates of both Q2 and GaN FET Q3 low turning them off. When voltage on the Q1 gate is driven low, its drain rises and raises the voltage on the Q2 gate turning it on that in turn forces current to flow out of the Q2 source to charge GaN FET Q3 gate capacitance turning it on. Diode D2 is reverse biased when Q1 drain voltage rises allowing Q2 to turn on by preventing R2 from being shunted around Q2. Diode D2 forward biases when Q1 turns on and, being a low resistance, allows high current to discharge Q3 gate capacitance to turn it off as fast as possible.

There needs to be resistance in series with the Q2 gate when current flows in both directions – into the gate when turning on and out of the gate when it turns off – otherwise deleterious ringing occurs due to circuit parasitics. R1 damps ringing when Q2 turns on. Usually, FET damping is provided by placing a resistor directly in series with the gate. However, with this circuit, resistor R2 not only damps ringing when Q2 turns off but also reduces current through R1 when Q1 turns on thereby reducing power dissipated in R1. Also, R2 is small enough in value that voltage across it does not prevent Q2 from turning off when it should.

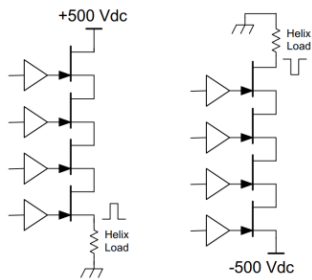

Fig. 2. Two 4-transistor switches to drive upper and lower helix of the kicker

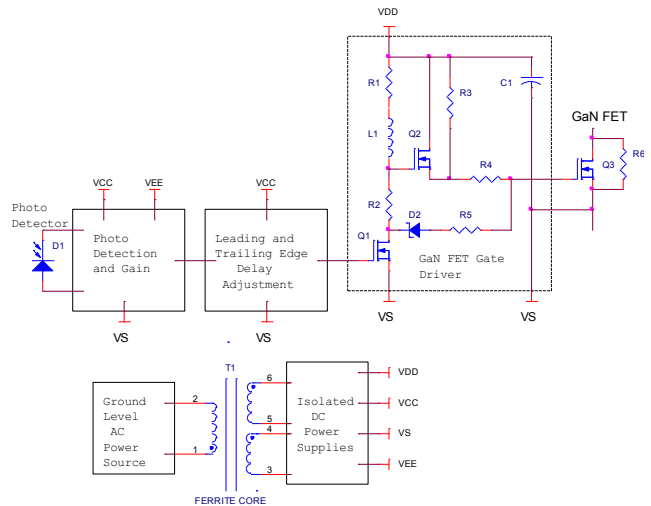

Fig. 3. Fermilab's gate driver for GaN transistors

Several factors contribute to the novel and power-efficient design of the GaN FET driver circuit and switching speed. One is that Q1 and Q2 were very uniquely chosen. They are GaN FETs that have very fast switching speed, are physically very small, have low gate to source and drain to source capacitance in relation to their drain current rating and their turn-on threshold voltage is lower than any other type of FET.

Transistors Q1 and Q2 are only 2 millimeters square in size that allows them to be physically located very close together and close to the Q3 gate. This minimizes the parasitic inductance along the current path through Q3 gate allowing R4 and R5 resistor values to be low enough to damp voltage ringing and still enable the Q3 gate to transition on and off in less than 2 ns.

The low capacitance values of Q1 and Q2 enables achieving the desired switching speed in several ways. The digital logic IC driving the gate of Q1, although having limited output drive current capability, can turn Q1 on and off in one nanosecond, because the Q1 gate capacitance is low. Also, a novel use of inductor L1 in series with R1 forms a two-pole network in combination with the sum of Q1 drain and Q2 gate capacitances. This network results in turning on Q2 with the speed as if a smaller value of R1 was used instead with L1 omitted. The resulting larger value of R1 reduces its power dissipation by about 30% compared with using a smaller R1 value without L1. The values of R1 and L1 are chosen for the two-pole network to be somewhat underdamped that results in the gate Q2 voltage to overshoot positive at turn on. However, because the Q1 and Q2 capacitances are as low as they are, the network's natural resonant frequency is high enough that the voltage settles back in time to not cause timing shifts if a turn on cycle occurs every 12 nanoseconds. This allows the multi-FET switch to achieve 80 MHz repetition switching rates. The picture of an assembled driver is shown in fig. 4

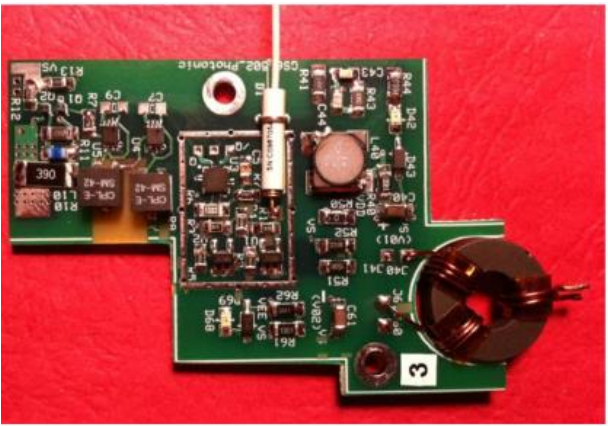

Fig. 4. Photo of Fermilab's GaN FET gate driver

## IV. THE PHOTONIC TRIGGER SYSTEM

The second problem to achieve high repetition rates is related to the delivery of fast and reliable triggers to gate drivers. In our previous publication [6] we used transformers to trigger the gate drivers. However, only optical triggering allowed us to achieve noise immunity as well as the switching rate required by PIP-II specifications. This photonics triggering system is shown in fig. 5. This system consists of a photonic transmitter and a photonic receiver separated by an optical splitter. The photonic transmitter provides an input to the optical splitter which provides a plurality of output signals to the photonic receivers. The photonic transmitter includes a laser diode that provides an input signal to a semiconductor optical amplifier (SOA). The SOA receives electrical pulses from a pulse generator. The SOA is essentially an optical shutter, because it is turned on and off very quickly to produce pulses of light. The SOA is output to the optical splitter. The photonic receiver includes a photodetector connected to an amplifier and comparator at the input to each driver circuit.

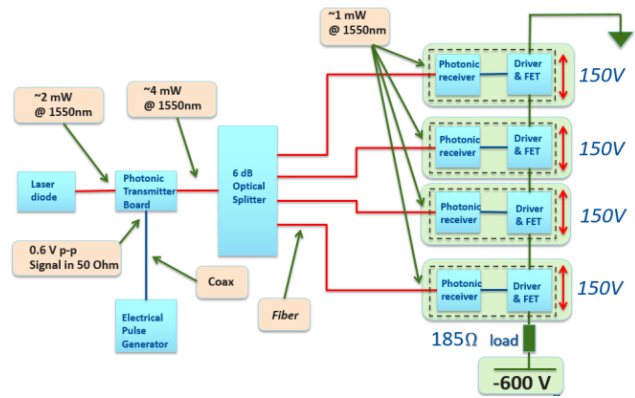

Fig. 5. Switch with the photonic trigger system

## V. TEST RESULTS

Figures 6 and 7 illustrate voltages on the kicker load, produced under different conditions. The 81.25 MHz waveform was only used for evaluation purposes. This waveform was chosen to demonstrate the capability to kick individually one bunch and pass individually one bunch, since bunches occur at 162.5 MHz repetition rate.

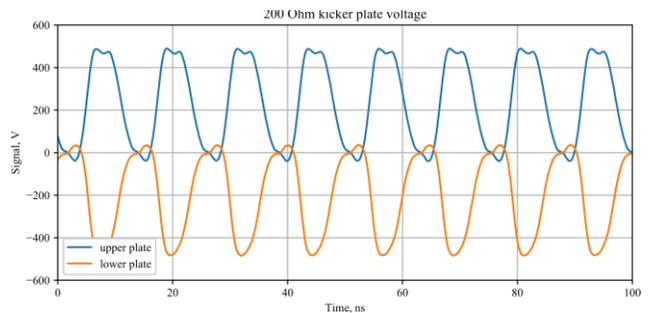

Fig. 6. 81.25 MHz switching, the actual burst lasts 12 microseconds.

One of the key features of this system is generating arbitrary pulse patterns as required by PIP-II parameters. The waveform shown in Fig. 7 is a short part of the required 0.55 ms arbitrary pattern burst during which the average switching frequency is 45 MHz. The total average switching rate of the system is currently limited to ~500 kHz. The main reason for this limitation is GaN transistors' overheating. It is not that it is too hot for the transistors, it is that the temperature rises to a point that changes the gate threshold voltage differently among the transistors and causes a timing mismatch. The limitation to ultimate repetition rates is power dissipation in the transistors. For example, our estimations show that ~35 MHz average switching rate will result in ~200 W/cm$^2$ heat flux density from each transistor in a four-transistor switch. This requires further R&D for adequate cooling. Such a cooling system must be done with completely dielectrically isolated components to absolutely minimize transistor parasitic capacitance to ground, to not compromise switching speed.

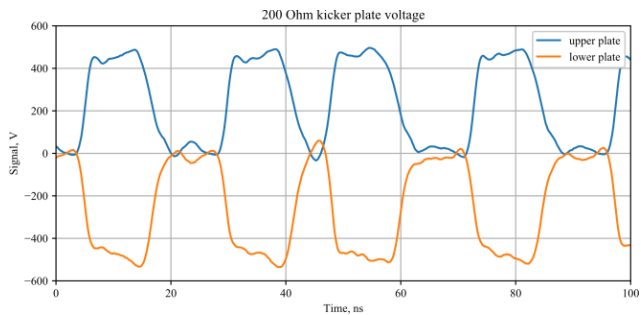

Fig. 7. Switching arbitrary pulse pattern.

## VI. CONCLUSION

We have demonstrated successful simultaneous control of four GaN transistors in series acting as a single switch with ~2 ns rise/fall time and high switching rates. To achieve it we developed a GaN gate driver circuit and photonic trigger system that assures precise control of on/off timing. The switch biased at 500 V operated at 81.25 MHz for 40 microseconds bursts and 45 MHz for 600 microsecond arbitrary bursts at 20 Hz repetition rate into a 185 Ohm load. Average switching rates are limited by cooling requirements. This topology is suitable for higher switching rates with development of adequate cooling.

## ACKNOWLEDGMENT

We would like to thank many people who helped us during several years of development and believed in our success, especially: Dan Wolff, Howie Pfeffer and Chris Jensen, for strategic and technical guidance and support; Sasha Shemyakin for persistent support while maintaining clear objectives; Brian Chase for considering a helical kicker structure; Alex Chen for kicker mechanical design and modeling a FET cooling option for CW operation; and Jeff Simmons for tireless assistance with assembly, testing, drafting and various technician support tasks.